\documentclass[aps,pra,twocolumn,amsmath,amssymb,showpacs]{revtex4}
\usepackage{graphicx}
\usepackage{epsfig}
\usepackage{dcolumn}
\usepackage{bm}

\usepackage{amssymb}
\usepackage{bm}
\usepackage{amsmath}

\begin{document}

\title{Ground state of a resonant two-qubit Rabi model in the ultrastrong coupling regime}
\author{Kelvin M. C. Lee and C. K. Law}
\affiliation{Department of Physics and Institute of Theoretical
Physics, The Chinese University of Hong Kong, Shatin, Hong Kong
Special Administrative Region, People's Republic of China}

\begin{abstract}
We consider a generalized Rabi model formed by two identical
qubits interacting with a common oscillator mode. In the near
resonance configuration where the oscillator frequency is close to
the transition frequency of the qubit, we determine the ground
state of the model approximately by using a variational method and
a transformation method. Both methods are shown to provide good
agreement with the exact numerical ground state for a range of
coupling strength in the ultrastrong regime. In addition, we
indicate how the accuracy of the approximation can be further
improved by using perturbation theory. We also examine the ground
state entanglement between the two qubits. By calculating the
negativity, we quantify the degree of entanglement as a function
of coupling strength.

\end{abstract}

 \maketitle

\section{Introduction}

Recently, there has been a revival of interest in studying the
behavior of a two-level system interacting with a quantum harmonic
oscillator beyond rotating wave approximation (RWA). One of the
motivations lies in the ultrastrong coupling regime recently
explored by experiments with artificial atoms and cavity photon
resonators \cite{Niemczyk,BSshift}, and microcavities embedding
doped quantum wells \cite{Gunter,Todorov}. In such a regime, the
qubit-oscillator interacting strength is comparable to the
oscillator frequency or the natural transition frequency of the
qubit, and theoretical investigations have found novel phenomena,
such as the asymmetry of vacuum Rabi-splitting \cite{Cao}, photon
blockade \cite{Ridolfo}, nonclassical states generation
\cite{Ashhab1}, superradiance transition \cite{Ashhab2}, and
collapse and revival dynamics \cite{Casanova}.

For a single qubit interacting with a quantum harmonic oscillator,
the system is described by the Rabi model \cite{Rabi} and it has
been studied extensively. While there exist analytical methods
\cite{Tur,Braak} to determine the ground state exactly,
approximation schemes are often employed because the approximate
ground states in their closed forms could be more physically
transparent for detailed investigations
\cite{Stolze,Chen,Irish,Gan,Zheng1,Zheng2,Yu,Liu,Hausinger,Hwang,Albert,Zhang}.

In this paper we investigate the ground state of a generalized
Rabi model in which a quantum harmonic oscillator interacts with
two identical qubits symmetrically. This is a special case of the
Tavis-Cummings model \cite{TC} beyond RWA. Recently Agarwal {\it
et al.} \cite{Eberly} have employed an adiabatic approximation to
determine the energy spectrum and entanglement dynamics in the
quasidegenerate situation with $\omega_A \ll \omega_c$, where
$\omega_A$ and $\omega_c$ are natural frequencies of the qubit and
the oscillator respectively. Our work here will focus on a
different situation with $\omega_A \approx \omega_c$. Such a near
resonance situation is more complicated because of the break down
of the adiabatic approximation.

To approach the problem, we employ two approximation schemes to
determine the ground state. In Sec.~II, we describe a variational
solution, in which the trial ground wave function is constructed
from a generalization of a single qubit problem
\cite{Stolze,Chen}. In Sec.~III, a transformation method is
introduced in order to simplify the Hamiltonian. Both methods turn
out to give solutions that have good agreement with the exact
numerical ground state. In particular the accuracy of the
transformation method can be further improved by combining with
perturbation theory.

An interesting feature in the system is the quantum entanglement
between the two qubits. In Sec.~IV, we shall characterize the
entanglement by using negativity. We shall see that there exists
entanglement even the system in the ground state. Such an
entanglement is due to counter-rotating terms involving virtual
photons. By using our approximate ground state, we quantify the
entanglement by negativity as a function of the coupling strength.

\section{The Hamiltonian and variational ground state}

The Hamiltonian of the model is given by
$(\hbar = 1)$
\begin{equation}
\label{eq:original hamiltonian}
H = \omega_{A} J_{x} + \omega_{c} a^{\dagger} a + g (a + a^{\dagger}) J_{z},
\end{equation}
where  $a$ and $a^{\dagger}$ are the annihilation and creation
operators of the cavity field mode of frequency $\omega_c$. The
$J$'s are spin-1 angular momentum operators. Physically, the
spin-1 system can be formed by two identical two-level atoms in
the triplet space. Then $\omega_{A}$ is the atomic transition
frequency, and $g$ denotes the collective atom-field coupling
strength. We note that the (antisymmetric) singlet state formed by
the two atoms does not couple to the field and it does not
constitute the ground state of the system.

\begin{figure*}[tbp]
\center
\includegraphics[width=6 in]{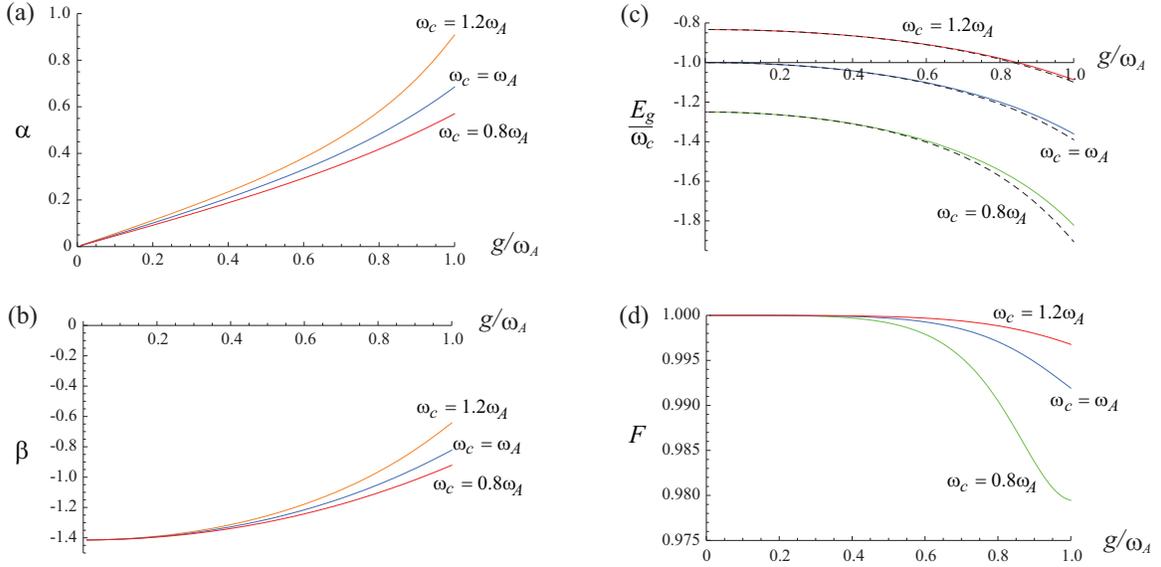}
\caption{(Color online) [(a)-(b)] Solution of the parameters
$\alpha$ and $\beta$ that minimize $\langle H \rangle$ as a
function of $g$ for various detunings. (c) The ground state energy
as a function of coupling strength ${g}/{\omega_{A}}$. The solid
line corresponds to $E_{v}$ from the variational method and the
dashed line corresponds to the exact numerical values $E_g$. (d)
The fidelity of the ground state obtained by the variational
method.} \label{transitions}
\end{figure*}

Our task in this paper is to determine the ground state energy $E_g$
and the ground state vector $|\psi_g \rangle$ of the system, which
are defined by:
\begin{equation}
H | \psi_g \rangle = E_g  | \psi_g \rangle.
\end{equation}
To this end, we let $|m \rangle_A $ be an eigenvector of $J_z$
(the quantum number $j=1$ is suppressed for convenience), i.e.,
$J_z |m \rangle_A = m |m \rangle_A$ with $m=0, \pm 1$. Then the
ground state of the system can be expressed as $|\psi_g \rangle =
\sum_{m=0, \pm 1} |f_m \rangle_F | m \rangle_A $, with  $|f_m
\rangle_F$ being the field state associated with the atomic state
$|m \rangle_A$.

To search for $|f_m \rangle$, we make use of variational method.
Specifically, we assume that $|f_m \rangle$ are coherent states so
that the trial ground state is in the form:
\begin{equation}
\left| {\psi_{v}} \right \rangle = \frac{1}{N} \left( {\left|
\alpha \right \rangle _{F} \left| {- 1} \right \rangle _{A} +
\beta \left| 0 \right \rangle _{F} \left| 0 \right \rangle _{A} +
\left| {- \alpha} \right \rangle _{F} \left| 1 \right \rangle
_{A}} \right).
\end{equation}
where $\left| \alpha \right \rangle _{F}$ is a coherent state of
the field with a real amplitude $\alpha$, $\beta$ is a variation
parameter, and $N^{2} = 2 + \beta^{2}$ is the normalization
constant. The expectation value of the Hamiltonian $\left \langle
H \right \rangle$ with respect to $\left| {\psi_{v}} \right
\rangle$ is given by,
\begin{equation}
\left \langle H \right \rangle = \frac{2}{2 + \beta^{2}} \left(
{\alpha^{2} \omega_{c} - 2 \alpha g + \sqrt{2} \beta \omega_{A}
e^{- \frac{\alpha^{2}}{2}}} \right).
\end{equation}
By varying the parameters $\alpha$ and $\beta$, $\left \langle H
\right \rangle$ is minimized at $\frac{\partial \left \langle H
\right \rangle}{\partial \alpha} = 0$ and $ \frac{\partial \left
\langle H \right \rangle}{\partial \beta} = 0$, which give the
conditions:
\begin{eqnarray}
\beta &=& \frac{\sqrt{2}}{\alpha \omega_{A}} \left( {\alpha
\omega_{c} - g} \right) e^{\frac{\alpha^{2}}{2}},
\\
g &=& \frac{\alpha^{2} \beta \omega_{c} - \left( {2 - \beta^{2}}
\right) \frac{\omega_{A}}{\sqrt{2}} e^{- \frac{\alpha ^{2}}{2}}}{2
\alpha \beta}.
\end{eqnarray}
The values of $\alpha$ and $\beta$ can be obtained by solving the
above equations numerically (Fig.~1a and b), and hence the
approximate ground state energy $E_{v} = \langle H \rangle$ can be
determined. We remark that there are two sets of $\alpha$ and
$\beta$ which satisfy the conditions given in Eqs.~(5) and (6),
but only the set with $\alpha < g/ \omega_c$ minimizes $\left
\langle H \right \rangle$. In addition, if the coupling strength
$g$ is smaller than $\omega_c + \omega_A$, then $\alpha$ and
$\beta$ take the following approximate forms:
\begin{eqnarray}
\alpha &\approx& \frac{g}{\omega_{A} + \omega_{c}},
\\
\beta &\approx& - \sqrt{2} + \frac{2\omega_{A} +
\omega_{c}}{\sqrt{2} \omega_{A} \left( {\omega_{A} + \omega_{c}}
\right)^{2}} g^{2}.
\end{eqnarray}
For the parameter used in Fig.~1, we see that Eqs.~(7) and (8) can
well capture the exact values of $\alpha$ and $\beta$ for small
$g$.

We have tested the accuracy of the variational method by comparing
the approximate ground state energy $E_{v}$ with the numerically
exact eigenvalues $E_g$ near the resonance situation with
$\omega_A \approx \omega_c$. The comparison is graphically shown
in Fig.~1c. We see that the variational solution can produce the
ground state energies very well agree with the exact values for
$g/\omega_A \leq 0.5$. For example, the error is about 0.1$\%$ at
$g= 0.5 ~\omega_A$ at the exact resonance. In fact, the variational
solution works in a wider range of $g$ when there is a positive
detuning ${\omega_{c}}- \omega_{A} > 0$. This is shown in the
curves $\omega_c= 1.2 ~\omega_A$ in which there is no visible
discrepancy for $g/\omega_A < 0.8$.

Another measure of the accuracy is the fidelity defined by inner
product between the exact numerical ground state vector and the
variational state vector, i.e., $F = \langle \psi_v | \psi_g
\rangle $. The results are shown in Fig.~1d, where we have $F
> 0.999$ for $g/\omega_A<0.5$ for the resonance case.

\section{Transformation method}

In this section, we describe an alternative method to obtain the
ground state of the system. Such a method is based on a unitary
transformation that simplifies the Hamiltonian to a form without
countering rotating terms approximately. In the case of the
(spin-half) Rabi model, the corresponding unitary transformation
has been discussed in Refs. \cite{Gan,Zheng1,Zheng2}. Here we
indicate how the method can be extended to our spin-1 model.

To begin with we consider a unitary transformation defined by $H'
= e^S H e^{-S}$. Here
\begin{equation}
S = \chi \left( {a^{\dagger} - a} \right) J_{z}
\end{equation}
with $\chi$ being a parameter to be determined. Under the unitary
transformation, $H'$ can be expressed as
\begin{equation}
\label{eq:transformed hamiltonian} H' = H'_{0} + H'_{1} + H'_{2},
\end{equation}
with
\begin{eqnarray}
\label{eq:separate transformed hamiltonian}  H'_{0} &=& \eta
\omega_{A} J_{x} - \left( {2 g \chi - \omega _{c} \chi^{2}}
\right) J_{z}^{2} + \omega_{c} a^{\dagger} a,
\\
 H'_{1} &=& \left( {g - \omega_c \chi} \right) \left( {a^{\dagger}
+ a} \right) J_{z} + i \eta \omega_{A} \chi \left( {a^{\dagger} -
a} \right) J_{y},
\\
 H'_{2} &=& \omega _{A} J_{x} \left\{ {\cosh \left[ {\chi \left(
{a^{\dagger} - a} \right)} \right] - \eta} \right\} \nonumber \\
&+& i \omega_{A} J_{y} \left\{ {\sinh \left[ {\chi \left(
{a^{\dagger} - a} \right)} \right] - \eta \chi \left( {a^{\dagger}
- a} \right)} \right\}.
\end{eqnarray}
Here $\eta = e^{- \frac{\chi^{2}}{2}}$ is defined. The $\cosh
\left[ {\chi \left( {a^{\dagger} - a} \right)} \right]$ and $\sinh
\left[ {\chi \left( {a^{\dagger} - a} \right)} \right]$ terms in
$H_2'$ has the leading expansion term:
\begin{eqnarray}
&& \cosh \left( {\chi \left( {a^{\dagger} - a} \right)} \right) =
\eta +  O\left( \chi^{2} \right),
\\
&& \sinh \left( {\chi \left( {a^{\dagger} - a} \right)} \right) =
\eta \chi \left( {a^{\dagger} - a} \right) + O \left( \chi^{3}
\right).
\end{eqnarray}
Therefore $H_2'$ is of order  $O(\chi^{2})$ describing higher
order (multi-photon) processes, and we shall neglect $H_2'$ as an
approximation, i.e., $H' \approx H_0' + H_1'$. This is the same
approximation procedure in the spin-half system
\cite{Gan,Zheng1,Zheng2}. However, the main difference here is the
presence of the $J_z^2 $ operator term in $H_0'$. This is in
contrast to the spin-half system in which the corresponding
$S_z^2$ term is just a constant.

To proceed, it is useful to express the atomic part of $ H_0' +
H_1'$ in the diagonal basis of  $\eta \omega_{A} J_{x} - \left( {2
g \chi - \omega _{c} \chi^{2}} \right) J_{z}^{2}$. Such an
operator appears in $H_0'$  and it describes a renormalized
three-level system. Specifically, we solve $[\eta \omega_{A} J_{x}
- \left( {2 g \chi - \omega _{c} \chi^{2}} \right) J_{z}^{2}] |\nu
\rangle= \varepsilon_{\nu} |\nu \rangle$ for eigenvectors $|\nu
\rangle$ and eigenvalues $\varepsilon_{\nu}$, with the index $\nu
= 0, \pm$ labelling the eigenvectors  such  that $\varepsilon_{+}
> \varepsilon_{0}
> \varepsilon_{-}$. It can be shown that $\varepsilon_{0}= - \mu
\eta \omega_{A} $, $\varepsilon_{\pm} = \eta \omega_{A} (- \mu \pm
\sqrt{4 + \mu^{2}}) /2$ with $\mu \equiv ({2 g \chi - \omega_{c}
\chi^{2}})/{\eta \omega_{A}}$, and the eigenvectors are:
\begin{eqnarray}
&& | \pm \rangle = \frac{1}{N_{\pm}}(|1 \rangle_A + \lambda_{\pm}
|0 \rangle_A + |-1 \rangle_A ),
\\
&& | 0 \rangle = \frac{1}{\sqrt{2}}(|1 \rangle_A  - |-1 \rangle_A),
\end{eqnarray}
where $\lambda_{\pm} = (\mu \pm \sqrt{4 + \mu^{2}}) /{\sqrt{2}}$
and $N_{\pm}^{2} = 2 + \lambda_{\pm}^{2}$ are the normalization
constants. In this way,
\begin{equation}
H' \approx  \omega_c a^\dag a I_A + \begin{pmatrix}
\varepsilon_{-}  & {c_{1} a + c_{2} a^{\dagger}} & 0 \\
{c_{2} a + c_{1} a^{\dagger}} & \varepsilon_{0} & {c_{3} a + c_{4} a^{\dagger}} \\
0 & {c_{4} a + c_{3} a^{\dagger}} & \varepsilon_{+}
\end{pmatrix},
\end{equation}
where $I_A$ is the $3 \times 3$ identity matrix operating on the
atomic subspace. The $c_j$ are some constants depending on the
parameter $\chi$. We see in Eq.~(18) that the terms involving
coefficients $c_1$ and $c_3$ are counter rotating, because they
describe virtual processes in which atomic excitation is
accompanied by the emission of a photon.

The main purpose of transforming $H$ into $H'$ is to eliminate
some of the counter rotating terms by using a suitable parameter
$\chi$.  Unlike the spin-half atom, it is not possible to find a
$\chi$ to remove all counter rotating terms. However, since we are
interested in the ground state, it is sufficient to find $\chi$
that makes $c_1=0$, because in this case the corresponding $H'$ in
Eq.~(18) has the eigenvector $|0 \rangle_F | - \rangle_ A$, which
is expected to be the ground state vector if the interaction is
not too strong. Returning to the original frame, we have the
approximation for the ground state energy:
\begin{equation}
E_g \approx  \varepsilon_{-} = - \frac{\eta \omega_{A} \left( {\mu
+ \sqrt{4 + \mu^{2}}} \right)}{2},
\end{equation}
and the ground state vector $\left| {\psi_{g}} \right \rangle
\approx \left| {\psi_{T}} \right \rangle$ (subscript $T$ for the
transformation method), where
\begin{eqnarray}
\left| {\psi_{T}} \right \rangle &=& e^{- S} |0 \rangle_F | -
\rangle_ A \nonumber
\\
&=& \frac{1}{N_{-}} \left( {\left| \chi \right \rangle _{F} \left|
{- 1} \right \rangle _{A} + \lambda_{-} \left| 0 \right \rangle
_{F} \left| 0 \right \rangle _{A} + \left| {- \chi} \right \rangle
_{F} \left| 1 \right \rangle _{A}} \right). \nonumber \\
\end{eqnarray}
Here $\left| \chi \right \rangle _{F}$ denotes the coherent state
of the field with the amplitude $\chi$. We see that this is
exactly the same form of the trial variational function given in
Eq.~(3). The value of $\chi$ is determined by the condition
$c_1=0$ which gives,
\begin{equation}
\lambda_- = \frac{\sqrt{2}}{\chi \omega_{A}} \left( {\chi
\omega_{c} - g} \right) e^{\frac{\chi^{2}}{2}}.
\end{equation}
Moreover, by $\lambda_{-} = \frac{\mu - \sqrt{4 +
\mu^{2}}}{\sqrt{2}}$ and $\mu = \frac{2 g \chi - \omega_{c}
\chi^{2}}{\eta \omega_{A}}$, we have
\begin{equation}
g = \frac{\chi^{2} \lambda_{-} \omega_{c} - \left( {2 -
\lambda_{-}^{2}} \right) \frac{\omega_{A}}{\sqrt{2}}
e^{-\frac{\chi^{2}}{2}}}{2 \chi \lambda_{-}}.
\end{equation}
Comparing Eqs.~(21) and (22) with Eqs.~(5) and (6) obtained in the
variational method, we can see that the transformation method and
the variational method give the same ground state, with $\chi$ and
$\lambda_{-}$ are equivalent to $\alpha$ and $\beta$ respectively,
i.e., we have $\varepsilon_{-}=E_{v}$, $|\psi_T \rangle = |\psi_v
\rangle$. Therefore the transformation procedure in this section
can be understood as a variation approach, in the sense that we
vary the Hamiltonian by the unitary transformation, instead of the
trial wave function.

However, the main advantage of the transformation method is that
the corrections can be identified explicitly. This is because
$H_2' \approx 0$ is the only approximation made in the above
derivation. Therefore the exact ground state involves higher order
processes described by $H_2'$, and this information cannot be
obtained by the simply varying the wave function as in Sec.~II.
Specifically, for $\chi <1 $, the leading contribution of $H_2'$
is
\begin{equation}
H_2' \approx \frac{\eta \chi^{2}\omega _{A} }{2}  J_{x}  \left(
{a^{\dagger 2} - 2 a^{\dagger} a + a^{2}} \right),
\end{equation}
which comes from the expansion of the $\cosh [ {\chi \left(
{a^{\dagger} - a} \right)}]$ term. By treating Eq.~(23) as a weak
perturbation, we can use the perturbation theory to find the
correction of the ground state energy (up to second order of
$H_2'$):
\begin{eqnarray}
&&  E_{g}   \approx  \varepsilon_{-} + \delta E,
\\
&& \delta E = - \frac{2 \chi^{4}}{N_{-}^{2}} \left[ {\frac{2
\varepsilon_{+}^{2}}{N_{-}^{2} \omega_{c}} +
\frac{\varepsilon_{0}^{2}}{N_{+}^{2} \left( {2 \omega_{c} +
\varepsilon_{+} - \varepsilon_{-}} \right)}} \right].
\end{eqnarray}
In Table~I, we indicate the performance of the approximation
scheme for the resonance case $\omega_A = \omega_c$. We see that
although $\varepsilon_{-}$ is already quite close to the exact
eigenvalue $E_g$, there is a significant improvement by including
the correction $\delta E$, by comparing the second and forth
columns in the table.

\begin{table}[h]
\begin{tabular}{llll}
\toprule
$~{g}/{\omega_{A}}~$ & $\ \ \ ~E_g/ \omega_A~$ & $\ \ \ ~\varepsilon_{-} / \omega_A~$ & $ \ ~(\varepsilon_{-} + \delta E)/\omega_A~$  \\
\colrule \ ~0.2~ & ~\ -1.01015~ & \ ~-1.01013~ & \ \ \ \ ~-1.01015~  \\
~ 0.4~ &
\ ~-1.04256~ & \ ~-1.04210~ & \ \ \ \ ~-1.04255~\\
\ ~0.6~ & \ ~-1.10404~ & \ ~-1.10137~ &\ \ \ \ ~-1.10403~ \\  \
~0.8~ &\ ~-1.20984~ & \ ~-1.19965~ & \ \ \ \ ~-1.20988~  \\
\ ~1.0~ & \ ~-1.38986~ & \ ~-1.36052~ & \ \ \ \ ~-1.39094~  \\
\ ~1.2~ & \ ~-1.68602~ & \ ~-1.62699~ &\ \ \ \  ~-1.68995~  \\
\botrule
\end{tabular}
\caption{A list of ground state energy obtained by various methods
at $\omega_c=\omega_A$. $E_g$, $\varepsilon_{-}$, and $
(\varepsilon_{-} + \delta E)/\omega_A$ are defined in the text.}
\label{tab1}
\end{table}

\section{Quantum entanglement between qubits}

In this section we examine the quantum entanglement between the
two qubits when the system is in the ground state. Negativity is
employed in order to quantify how the two qubits are entangled in
the ultrastrong coupling regime. The negativity is defined as
\cite{Vidal}
\begin{eqnarray}
\mathcal{N} \left( \rho_{A} \right) \equiv \frac{\left \| \rho_{A}^{T_{A}} \right \|_{1} - 1}{2},
\end{eqnarray}
where $\left \| \rho_{A}^{T_{A}} \right \|_{1}$ is the trace norm
of $\rho_{A}^{T_{A}}$, $\rho_{A}^{T_{A}}$ is the partial transpose
of the reduced density matrix $\rho_{A} = {\rm Tr}_{F} \left(
{\rho} \right)$ and $\rho = \left| \psi_{g} \right \rangle \left
\langle \psi_{g} \right|$. Alternatively, Eq.~(26) can be
calculated by the absolute value of the sum of the negative
eigenvalues of $\rho^{T_{A}}_{A}$.

Using the approximate ground state vector $\left| \psi_{v} \right
\rangle$ in Eq.~(3), the corresponding reduced density matrix
$\rho_{A}$ can be expressed as
\begin{eqnarray}
\rho_{A} = \frac{1}{2 N^{2}} \begin{pmatrix}
\rho_{11} & 0 & 0 & \rho_{14} \\
0 & \rho_{22} & \rho_{23} & 0 \\
0 & \rho_{32} & \rho_{33} & 0 \\
\rho_{41} & 0 & 0 & \rho_{44}
\end{pmatrix},
\end{eqnarray}
where
\begin{eqnarray}
&&\rho_{11} = 1 + \beta^{2} + 2 \sqrt{2} \beta e^{- \frac{\alpha^{2}}{2}} + e^{- 2 \alpha^{2}},
\\
&&\rho_{14} = \rho_{41} = 1 - \beta^{2} + e^{- 2 \alpha^{2}},
\\
&&\rho_{22} = \rho_{23} = \rho_{32} = \rho_{33} = 1 - e^{- 2 \alpha^{2}},
\\
&&\rho_{44} = 1 + \beta^{2} - 2 \sqrt{2} \beta e^{- \frac{\alpha^{2}}{2}} + e^{- 2 \alpha^{2}},
\end{eqnarray}
and the order of the columns and rows is $\left| {e e} \right
\rangle$, $\left| {e g} \right \rangle$, $\left| {g e} \right
\rangle$ and $\left| {g g} \right \rangle$ which correspond to the
bare atomic level of the two qubits. The four eigenvalues of
$\rho^{T_{A}}_{A}$ can be obtained analytically and only one of
them is negative. Thus, the expression of the approximate
negativity $\mathcal{N}$ is
\begin{eqnarray}
\mathcal{N} = max \left \{ \frac{2 e^{- 2 \alpha^{2}} - \beta^{2}}{2 \left( {2 + \beta^{2}} \right)} ~, ~0 \right \}.
\end{eqnarray}
In the limit $g/\omega_A \ll 1$, $\alpha$ and $\beta$ are
approximated by Eqs. (7) and (8), and this gives
\begin{eqnarray}
\mathcal{N} \approx \frac{\omega_{c}}{4 \omega_{A} \left( {\omega_{A} + \omega_{c}} \right)^{2}} g^{2}.
\end{eqnarray}
Therefore the degree of entanglement increases with $g^2$ when
$g/\omega_A \ll 1$. As a remark, we have also calculated the
corresponding concurrence and it is simply twice of $\mathcal{N}$
calculated in the Eq.~(32).

We have computed the negativity from the numerically exact ground
state for the resonance case, and it is plotted in Fig.~2 with a
dashed line. We see that the negativity is well approximated by
Eq.~(32) (solid line) for $g/\omega_A <0.5$. In particular, the
quadratic dependence of $g$ is captured by Eq.~(33) at small $g$.
There is a greater discrepancy when $g/\omega_A
> 0.5$, which is consistent with fidelity $F$ behavior shown in
Fig.~1d. Although Fig.~2 does not show the case of $g/\omega_A
> 1$, the exact negativity drops as $g/\omega_A$ further increases after
it reaches its maximum at $g \approx \omega_{A}$. When $g \approx
2.6 ~\omega_{A}$, the negativity would drop to zero and never
increase again. In other words, the two qubits become disentangled
if $g$ is sufficiently large.

\begin{figure}[htb]
\centering
\includegraphics[width = 0.45\textwidth]{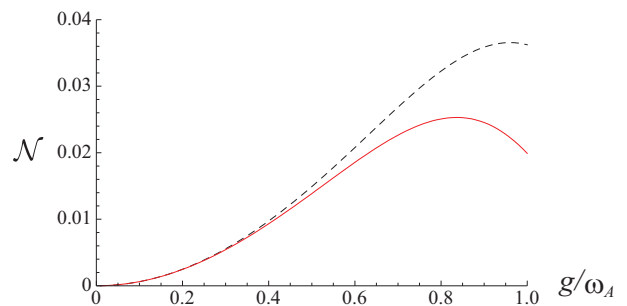}
\caption{(Color online) The negativity $\mathcal{N}$ as a function
of coupling strength ${g}/{\omega_{A}}$ for the $\omega_{c} =
\omega_{A}$ case. The solid line (dashed line) corresponds to the
negativity of the approximate (exact) ground state.}
\label{fig:negativity}
\end{figure}

\section{Conclusion}

To conclude, we have shown that the ground state of the Rabi model
formed by two identical qubits in the near resonance configuration
can be determined approximately by using a variational method and
a transformation method, and the results are in good agreement
with exact numerical calculations when $g$ is a significant
fraction of $\omega_A$ (up to $g/\omega_A =0.5$). For instance,
the error of the variational ground state energy for the exact
resonance case is about $0.1\%$ at $g = 0.5 ~\omega_A$, and the
error can be greatly reduced to 0.0004$\%$ by further using second
order perturbation theory. The key advantage of our methods is
that the analytical form of the ground state can be captured
approximately, which is useful for analyzing the ground state
properties. Specifically, we have examined the quantum
entanglement between the two qubits. Such an entanglement is a
consequence of counter rotating terms in the Hamiltonian, because
if such terms are dropped by RWA, then the two qubits are simply
disentangled in the ground state. Using our approximate ground
state vector, we are able to determine the negativity as a
function of system parameters for $g/\omega_A$ up to 0.5.

\begin{acknowledgments}
This work is partially supported by a grant from the Research
Grants Council of Hong Kong, Special Administrative Region of
China (Project No.~CUHK401812).
\end{acknowledgments}

\end{document}